\documentclass[aps,10pt,pra,twocolumn,groupedaddress,floatfix, superscriptaddress,showpacs,showkeys,amsfonts]{revtex4-1}
\usepackage{amsfonts}
\usepackage{amsmath}
\usepackage{amssymb, setspace}
\usepackage{graphics,graphicx}
\graphicspath{{./}}
\usepackage{epsf}
\usepackage{subfigure}
\usepackage[%
  colorlinks=true,
  urlcolor=blue,
  linkcolor=blue,
  citecolor=blue
]{hyperref}
\usepackage[all]{hypcap}
\usepackage{color}
\usepackage[utf8]{inputenc}
\usepackage{dsfont}
\usepackage{enumitem}

\newcommand{\be}{\begin{equation}}
\newcommand{\ee}{\end{equation}}
\newcommand{\bea}{\begin{eqnarray}}
\newcommand{\eea}{\end{eqnarray}}
\newcommand{\eq}[1]{Eq.~\eqref{#1}}
\newcommand{\eqs}[1]{Eqs.~\eqref{#1}}
\newcommand{\eqss}[2]{Eqs.~\eqref{#1}-\eqref{#2}}
\newcommand{\seq}[1]{Sec.~\ref{#1}}

\newcommand{\app}[1]{App.~\ref{#1}}
\newcommand{\fig}[1]{Fig.~\ref{#1}}
\newcommand{\figs}[2]{Figs.~\ref{#1}-\ref{#2}}
\newcommand{\bem}{\begin{multline}}
\newcommand{\eem}{\end{multline}}

\newcommand{\tr}[1]{{\rm tr}\left\{#1\right\}}

\graphicspath{ {figures/} }

\begin{document}

\title{Experimental Bayesian estimation of quantum state preparation, measurement, and gate errors in multi-qubit devices}

\author{Haggai Landa}\email{haggai.landa@ibm.com}
\author{Dekel Meirom}
\affiliation{IBM Quantum, IBM Research -- Haifa, Haifa University Campus, Mount Carmel, Haifa 31905, Israel}
\author{Naoki Kanazawa}
\affiliation{IBM Quantum, IBM Research -- Tokyo, 19-21 Nihonbashi Hakozaki-cho, Chuo-ku, Tokyo, 103-8510, Japan}
\author{Mattias Fitzpatrick}
\author{Christopher J. Wood}
\affiliation{IBM Quantum, T.J.~Watson Research Center, Yorktown Heights, NY 10598, USA}

\begin{abstract}

We introduce a Bayesian method for the estimation of single qubit errors in quantum devices, and use it to characterize these errors on three superconducting 27-qubit devices. We self-consistently estimate up to seven parameters of each qubit's state preparation, readout, and gate errors, analyze the stability of these errors as a function of time,
and demonstrate easily implemented approaches for mitigating different errors before a quantum computation experiment. On the investigated devices we find non-negligible qubit reset errors that  cannot be parametrized as a diagonal mixed state, but manifest as a coherent phase in a superposition with a small contribution from the qubit's excited state. We are able to mitigate such errors by applying pre-rotations on the initialized qubits, which we demonstrate with multi-qubit entangled states. Our results show that Bayesian estimation can resolve small parameters -- including those pertaining to quantum gate errors -- with a high relative accuracy, at a lower measurement cost as compared with standard characterization approaches.
\end{abstract}

\maketitle

\section{Introduction}\label{Sec:Intro}

Quantum computers experience a variety of errors that limit the ability of near-term quantum devices to perform arbitrary computations in the absence of fault-tolerant quantum error correction. Characterizing the performance of these devices is a challenge of significant importance for their use with near-term applications~\cite{corcoles2020challanges}. Recently a variety of error mitigation techniques have been proposed to reduce the effects of errors in current devices, with the goal of achieving useful quantum computation on noisy near-term quantum computers~\cite{kandala2019error,sun2021mitigating}. Many of these techniques focus on mitigating measurement errors through classical post-processing of measurement data based on an accurately characterized model of errors in measurement~\cite{chen2019detector,bravyi2021mitigating, Maciejewski2020mitigation,barron2020measurement, Geller_2021}. However, characterization of measurement (readout) errors in isolation of preparation or gate errors in particular can be challenging. Other approaches aim to mitigate specific kinds of gate errors by applying post-processing of data from additional modified experiments to average or extrapolate away the effect of errors~\cite{temme2017error,endo2018practical,giurgica2020digital,berg2020model}, where typically errors must be characterized to confirm they are a good match to the model. 

Different errors require a variety of techniques to characterize and mitigate and it is often useful to classify them into distinct types of preparation, measurement, and gate errors. In the circuit model a typical quantum experiment consists of (i) Preparation (reset) to initialize specific qubits in a fixed fiducial state,(ii) applying unitary gates to evolve the state, and (iii) measurement of qubits  in a fixed basis to extract state probabilities.
Errors in steps (i) and (iii) -- so called state preparation and measurement (SPAM) errors -- are typically treated separately from the gate errors (although the distinction could depend on the context). 

For gate errors one widely used technique is quantum process tomography (QPT), which aims to estimate the process matrix of a noisy gate from accessible measurements of a set of known input states~\cite{chuang1997prescription,fiuravsek2001maximum, shabani2011efficient,flammia2012quantum}. This technique is however susceptible to SPAM errors which are difficult to distinguish from the gate errors it aims to characterize.
Gateset tomography (GST) is a generalization that simultaneously and self-consistently characterizes the performance of preparation, measurement, and gates of a small generating gateset \cite{merkel2013self,blume2013robust,blume2017demonstration}. 
Other commonly used methods for estimating average gate errors in gatesets are randomized benchmarking (RB) and many related randomized protocols~\cite{kimmel2014robust,johnson2015demonstration,
proctor2019direct,erhard2019characterizing,
harper2020efficient}. These decouple the effect of SPAM errors by fitting decay parameters to an exponential noise model as a function of sequence lengths and are used extensively in current experiments~\cite{mccormick2017experimental,keith2018joint, ferrie2014self,chapman2016experimental, granade2017practical, knee2018quantum}.
Another approach is based on constructing experiment protocols that amplify and isolate certain parameters in the results, while making the signal independent of the values of other (unknown) parameters, in particular SPAM errors. Such sequences are often referred to as error amplification \cite{sheldon2016characterizing}, and are in common use.

Works that employ Bayesian estimation techniques in the context of quantum parameter estimation have been proposed for quantum state tomography (QST) \cite{blume2010optimal}, Hamiltonian learning \cite{wang2017experimental,evans2019scalable, Bairey_2019}, and randomized benchmarking \cite{granade2015accelerated}. Alternative QST methods include self-guided tomography \cite{PhysRevLett.113.190404, PhysRevLett.117.040402, PhysRevLett.126.100402}. Experimental estimates of Hamiltonian parameters in a system of nitrogen vacancy in a diamond have been obtained \cite{hincks2018hamiltonian} using Monte Carlo simulations and also involved what is known as online Bayesian experiment design, where experiment steps are modified online as data is processed.

In this work, we present a characterization of gate and SPAM errors using an analytic Bayesian approach. We consider preparation errors as resulting from a noisy process that prepares an initial density matrix different from the intended pure state, and measurement errors as a positive operator-valued measure (POVM) that differs from an ideal projective measurement of the diagonal elements of a density matrix.
We distinguish between four type of SPAM errors; classical vs.~quantum (coherent) preparation errors, and classical vs.~quantum measurement (readout) errors. We analyze both the general case and special cases of classical preparation errors, quantum preparation errors, and classical readout errors. Methods for the characterization and mitigation of classical SPAM errors are extensively researched and are relatively straightforward. In contrast, the estimation and mitigation of quantum preparation errors is less developed, and constitutes one of the main contributions of the current work. 

We develop the theory in detail for single-qubit errors, focusing on errors that have been observed in our experiments, while emphasizing that the generalization of the approach to parametrize multi-qubit errors is formally straightforward. Our method estimates the probability distribution of the parameters of interest, given a prior distribution (that we take as uniform in a relevant domain of the parameter space) and an assumed model. We take advantage of closed-form relations between the probability of experimental measurements and the parameters of the model to cope with the main downside of Bayesian estimation: the computational burden. The advantage of the Bayesian approach is that the full information in the experiment data is utilized to self-consistently characterize all parameters simultaneously. The error estimates then immediately follow and the covariances of the parameters are accounted for. In addition, physical constraints are enforced \emph{ab inito} (into the parameter priors), avoiding typical complications of fitting methods that may give unphysical results. This allows for a reliable characterization of even small parameters and fully exploits the information about all parameters contained in the experiments -- significantly reducing the required experimental resources. 

These techniques were applied to characterize single-qubit SPAM errors in the 27-qubit IBM Quantum devices \emph{ibmq\_paris}, \emph{ibmq\_sydney},  and  \emph{ibmq\_toronto}, accessible via the cloud using Qiskit \cite{Qiskit}. An open-source code repository \cite{quantum-bayesian-estimation-repo} implements our Bayesian estimation approach and can be used to run similar estimation and research experiments. We observe that SPAM errors are mostly uncorrelated between qubits in the devices studied. Hence, the characterization can be done independently (and in parallel) for all qubits of a device, or in two disconnected groups of qubits, a common practice.  Quantum (coherent) preparation errors are found to be stable in time, suggesting that these coherent errors can be corrected {\it a-priori}. 

There are two possible ways for such a correction -- either the state preparation can be recalibrated to account for and reduce such errors, or the characterization data from one experiment can be used to perform a pre-rotation of each qubit at the beginning of a subsequent experiment. The calibration of the state preparation is beyond the scope of the current work and is beyond the capabilities of most cloud-based users so we focus on the second possibility of rotating the qubits before an experiment, demonstrating its usefulness by mitigating expectation values after creating (noisy) multi-qubit Greenberger–Horne–Zeilinger (GHZ) entangled states of up to 7 qubits. In addition, we estimate small gate errors and study the precision with which the Bayesian approach can be used to estimate and correct those, consistently finding a significant reduction in the required experiment measurements for the same level of precision as compared with standard characterization and error amplification techniques.

\section{Bayesian Estimation}\label{Sec:Bayesian}

\subsection{POVM measurement errors}\label{Sec:POVMErrors}

For a single qubit it is convenient to write the qubit's density matrix $\rho$ using the Bloch representation,
\be \rho = \frac{1}{2}\left(\mathds{1}+\vec{r}\cdot\vec{\sigma}\right)\equiv \frac{1}{2} \sum_{\mu} r_{\mu}\sigma^\mu,\quad \mu \in\{x,y,z,0\},\label{Eq:rhomu}\ee
where we have defined 
\be  r_0\equiv 1,\qquad \sigma^0\equiv \mathds{1},\ee
where $\mathds{1}$ is the identity matrix and $\sigma^a$ with $a\in\{x,y,z\}$ are the Pauli matrices. The vector of coefficients $\vec{r}$ is the Bloch vector representing the qubit's state within the Bloch sphere ($|\vec{r}|\leq 1$), since
\be\langle \sigma^{a}\rangle = {\rm tr}\{\rho\sigma^{a}\} = r_a.\ee 

We consider a measurement process in the computational basis (the basis of eigenstates of $\sigma^z$). A general two-outcome measurement can be modeled using a positive operator valued measure (POVM) $\mathcal{E}$,
\be \mathcal{E}=\{\Pi,\mathds{1}- \Pi\}, \text{where} \qquad \Pi \ge 0,\qquad \mathds{1}-  \Pi \ge 0. \label{Eq:POVMconstraints1Q} \ee 
The POVM element corresponding to the result 0 is expanded using four parameters $\pi_\mu$ as
\be \Pi = \sum_\mu \pi_{\mu}\sigma^\mu,\qquad \mu \in\{x,y,z,0\}\label{Eq:Pi}\ee
with some constraints on the parameters that are discussed below.  For the ideal measurement, $\Pi =|0\rangle\langle 0| =\frac{1}{2} \left( \mathds{1} +\sigma^z\right)$ is a projector on the state $|0\rangle$, with $\pi_0=\pi_z=1/2$, and $\mathds{1}-  \Pi $ is the projector on $|1\rangle$.
The probability of measuring the result 0 (that corresponds to the POVM element $\Pi$), given the state $\rho$ is found using \eq{Eq:Pi} and \eq{Eq:rhomu} where,
\be p\equiv p(0|\rho) = \tr{\rho \Pi}= 
 \sum_{\mu } r_\mu \pi_\mu.\label{Eq:p0r}\ee 

The single-qubit POVM parameters $\pi_\mu$ defined in \eq{Eq:Pi} generalize the classical measurement error parameters that can be defined as the conditional probabilities
 \be \epsilon_0 = p\left(1\, |\, \rho =|0\rangle\langle 0| \,\right) = 1 -\left( \pi_0 + \pi_z \right),\label{Eq:epsilon0} \ee
 \be \epsilon_1 = p\left( 0\, |\, \rho = |1\rangle \langle 1|\rangle \, \right)= \pi_0 - \pi_z.\label{Eq:epsilon1}\ee
Here $\epsilon_0$ is the probability of measuring 1 when the qubit is in the state $|0\rangle$, and  $\epsilon_1$ is the probability of measuring 0 when the qubit is in the state $|1\rangle$. We refer to these readout errors as the classical ones, since they correspond to classical bit flip errors.

The nonclassical measurement errors are given by the coefficients $\pi_x$ and $\pi_y$,
 which can be understood by considering the following conditional probabilities in the qubit's $x-y$ plane:
 \be p\left(0\, |\, \rho =|+\rangle\langle +| \,\right) = \pi_0 + \pi_x, \label{Eq:p_plus}\ee
 \be p\left( 0\, |\, \rho = \left| i\right\rangle \langle  i| \, \right)= \pi_0 + \pi_y,\label{Eq:p_i}\ee
 with $|+\rangle= ( |0\rangle+ |1\rangle)/\sqrt{2}$  and $|i\rangle= ( |0\rangle+i |1\rangle)/\sqrt{2}$. 
The probabilities in \eqss{Eq:p_plus}{Eq:p_i} correspond to measuring 0 when the qubit is aligned with the positive $x$ and $y$ axes, respectively. If the measurement process can be described using only the errors of \eqss{Eq:epsilon0}{Eq:epsilon1}, a qubit in the state $|+\rangle$ would collapse to either $|0\rangle$ or $|1\rangle$ with equal probability during the measurement, and the measurement error would be the arithmetic mean of the two errors, i.e., $\pi_0$. Thus, nonzero values for $\pi_x$ and $\pi_y$ represent nonclassical errors, sensitive to the qubit state being a superposition with a coherent phase, rather than a mixed state.

In \app{Sec:POVM4} we write the constraints on the POVM elements that stem from the positivity conditions in \eq{Eq:POVMconstraints1Q} in terms of the parameters, and we further discuss characterization of POVM errors in the case when both the initialization and the gate errors can be neglected, a scenario that lends itself to a straightforward analytic estimation. However, when there are initialization errors or gate errors in addition to measurement errors, the relations used in \app{Sec:POVM4} cannot be used.

\subsection{SPAM errors}\label{Sec:SPAMErrors}

Equation \eqref{Eq:p0r} contains seven independent parameters that describe the qubit state and the measurement POVM. When $\vec{r}$ is the initial state, these are the SPAM error parameters within this framework, and we introduce the notation
\be \vec{r}(t=0)\equiv \{x_0, y_0, z_0\},\ee
parametrizing the initial Bloch vector.
 The ideal initial state is $|0\rangle$, for which $z_0= 1$. In many cases, the only initialization error that is considered corresponds to a classical probability of the qubit being in the  $|1\rangle$ state. Then $\rho$ takes the form of a mixed diagonal state, with only $z_0$ being nonzero, and smaller than 1. With the $x_0$ and $y_0$ components, also a coherent (quantum) initialization error expressing an initial superposition of the qubit basis states (i.e., a rotated Bloch vector) can be modeled.
As discussed further in \seq{Sec:GateErrors}, gate errors can also be accounted for within the Bayesian approach [with \eq{Eq:p0r} generalized accordingly]. However, since in the current experiments these are found to be mostly relatively small (on the order of the estimation error bars, as argued in more detail and shown in \seq{Sec:BayesianResults}), we neglect gate errors at first, focusing in the current subsection on SPAM errors using an experiment consisting of six different gates followed by measurement, specifically
\be \left\{\mathds{1},X,X_{-90},X_{90},Y_{90},Y_{-90}\right\},\label{Eq:Calibration0}\ee
where $X=\sigma^x$, and $X_{\theta}$ ($Y_{\theta}$) correspond to a rotation of angle $\theta$ about the $x$ ($y$) axis.

The probabilities for measuring 0 in each of the six experiments using the gates in \eq{Eq:Calibration0} are given, respectively, by 
\be f_{\pm z} = \pi_0 +\pi_x x_0 \pm\pi_y y_0 \pm \pi_z z_0,\label{Eq:fpmz}\ee
\be f_{\pm y} = \pi_0\pm \pi_y z_0 +\pi_x x_0 \mp \pi_z y_0,\label{Eq:fpmy} \ee
\be f_{\pm x} = \pi_0 \pm \pi_x z_0 \mp\pi_z x_0 +\pi_y y_0.\label{Eq:fpmx}\ee
There are six independent relations in \eqss{Eq:fpmz}{Eq:fpmx}, and hence up to six parameters can be uniquely estimated. This is a known problem of SPAM error characterization (the parameters are sometimes said to be ``confounded''). We are not aware of a solution that the Bayesian method can offer specifically to this issue, and if all seven parameters have to be estimated together, further information beyond the gates considered above must be considered.

Considering the physics of the qubit initialization and readout, the main question is whether there are quantum readout errors (whence $\pi_x$ and $\pi_y$ cannot be neglected), and whether there are quantum initialization errors ($x_0$ and $y_0$ cannot be neglected). One can consider the current study as starting with an ansatz that retains the parameters $x_0$ and $y_0$ in the analysis, while $\pi_x$ and $\pi_y$ are neglected (set to 0). The results of  \seq{Sec:SPAMMitigationResults} then constitute a justification of this assumption {\it a posteriori}, since the active mitigation of $x_0$ and $y_0$ by a pre-rotation of each qubit works systematically. An analysis of the unconditional reset scheme implemented on IBM Quantum devices strengthens this conclusion and motivates the physical mechanism behind it. The essence of the unconditional reset pulse \cite{PhysRevApplied.10.044030} is a pumping of the qubit amplitude from the state $|1\rangle$ to a higher state (call it $|2\rangle$), followed by a resonant pulse inducing a transition from the state $|2\rangle$ to $|0\rangle$ accompanied by a creation of a photon in the coupled readout resonator, which then rapidly decays. This coupling is the leading order effective interaction induced by the reset driving, and is derived by adiabatically eliminating the off-resonant level $|1\rangle$. When the ramping up and down of the reset pulse is not adiabatic enough, a residual (coherent) population transfer into the $|1\rangle$ state is possible.

\subsection{Parametrization of gate errors}\label{Sec:GateErrors}

In addition to SPAM errors, the Bayesian framework described here can naturally incorporate gate errors, especially those which can be modeled analytically in closed form.
 As a simplified model for gate errors, we consider one incoherent error parameter, and one coherent gate error.
Typical qubit coherent errors include an amplitude error (of the driving pulse), a frequency error (in the estimated qubit frequency that determines the drive frequency), and angle error between $X$ and $Y$ rotations for the single qubit gate model.
 On the IBM Quantum devices, the angle offset of the $X$ rotation axis from the $Y$ axis is considered to be negligibly smaller as compared with other errors since those devices provide single-qubit basis gates consisting of $Z$ rotation and $X_{90}$ gates. Namely, the $Y$ rotation gate can be composed of  virtual-$Z$ gate \cite{mckay2017efficient} with exactly $\pi/2$ phase offset. Frequency errors are also typically well suppressed with the single-junction transmon qubits utilized in the IBM Quantum devices \cite{burnett2019decoherence}. 

Therefore, to account for coherent rotation errors, we assume that the $X_{\pm  90}, Y_{\pm 90}$ gates perform a rotation of the Bloch vector with one angle parametrizing deviations from an ideal gate. The rotation error angle $\theta$ is identical for all gates with just the axis of rotation and direction redefined for each gate. An extension of the model to arbitrary rotation errors is considered briefly in \app{App:Overrotations}.
 
 Repeated application of gates is necessary to amplify  small gate errors (as is typically the situation).
The result of repeated application of one of these gates $j$ times is calculated by taking the corresponding power of the rotation matrix and applying it to the initial state, meaning that 
\be (X_{\pm 90})^j =  \left(\begin{array}{ccc}
1 & 0 & 0 \\
 0 & \cos j(\frac{\pi}{2} + \theta) &  \mp \sin j(\frac{\pi}{2} + \theta) \\
0 & \pm \sin j(\frac{\pi}{2} + \theta) &  \cos j(\frac{\pi}{2} + \theta) \end{array}\right),\label{Eq:X90rotation}
\ee
and
\be (Y_{\pm 90})^j =  \left(\begin{array}{ccc}
\cos j(\frac{\pi}{2} + \theta) & 0 & \pm \sin j(\frac{\pi}{2} + \theta) \\
0 & 1 & 0 \\
\mp \sin j(\frac{\pi}{2} + \theta) & 0 & \cos j(\frac{\pi}{2} + \theta) \end{array}\right).\label{Eq:Y90rotation}
\ee
We note that the matrix power has to be computed analytically in order to allow the analytic calculation of the probabilities [App.~\ref{App:Overrotations} lists a few explicit calculations based on \eqss{Eq:X90rotation}{Eq:Y90rotation}].
 
 A uniform depolarizing error parameter $\varepsilon$ that acts on the density matrix during the operation of a gate repeated $j$ times can be modeled by the replacement
 \be \rho \to (1-\varepsilon)^j\rho +[ 1 - (1-\varepsilon)^j] \mathds{1},\label{Eq:depolarizing}\ee
 or, equivalently, $\vec{r}\to (1-\varepsilon)^j \vec{r}$. This single error parameter constitutes an approximate averaging of energy relaxation and dephasing errors (quantified using the characteristic times $T_1$ and $T_2$ \cite{krantz2019quantum}), that act on different qubit components within the Bloch sphere during the operation of the gate.
With gate errors, the probabilities in \eqss{Eq:fpmz}{Eq:fpmx} should not be used directly (as the gate errors were neglected there), but rather the computation proceeds by applying the gates to the initial state, and plugging the resulting Bloch vector into \eq{Eq:p0r} to obtain the final probabilities.

\subsection{Bayesian estimation}\label{Sec:BayesianIntegrals}

In the experiment, we prepare a fixed state $\rho$ and measure it for a total of $n$ shots. This process constitutes a sampling of a random variable, with the number of shots $s_0$ where the result $0$ is drawn from a binomial distribution with the parameter $p\equiv p(0|\rho)$ of \eq{Eq:p0r} is given by
\be s_0 \sim \mathcal{B}(n,p), \quad p(s_0=k) =\left(\substack{n\\k}\right) p^k(1-p)^{n-k}.\label{Eq:Binomial}\ee
We can estimate $p$ using the maximum likelihood estimator, 
 \be \hat p = s_0/n\label{Eq:hat_p_binomial}. \ee
 
From the experimental estimates of the six probabilities ($f_{\pm \mu}$), we cannot directly derive the distribution of the nonlinearly-dependent parameters of interest and therefore resort to a Bayesian approach.
In the following we denote the set of the unknown parameters by
\be \lambda=\{\lambda^0,...,\lambda^\alpha,...\}. \label{Eq:alpha1}\ee
In this notation, $\lambda$ is a vector,  $\lambda^\alpha$ is the $\alpha$'th component of this vector, while $\lambda_i$ and $\lambda_j$ will be used to denote two different vectors.

We define $p_e(\lambda)$ as the distribution of interest for the vector of parameters $\lambda$ given the set of observed measurements $\{m\}_e$
\be p_e(\lambda)  \equiv p(\lambda |\{m\}_e) = \frac{\tilde{p}_e(\lambda)}{\int \tilde{p}_e(\lambda) d\lambda },\label{Eq:Bayes1}\ee
where $p_e$ is normalized by integration of $\tilde{p}_e(\lambda)$ -- the unnormalized posterior density which is given by the Bayes theorem: 
\be \tilde{p}_e (\lambda) = p(\{m\}_e |\lambda) p_0(\lambda) .\label{Eq:Bayes0}\ee
Here, $p(\{m\}_e|\lambda)$ is the likelihood of the observed experiment (the probability of observing $\{m\}_e$ given $\lambda$), and $p_0(\lambda)$ is the prior distribution for the parameters, which we take as a uniform distribution over a bounded domain,
\be p_0(\lambda) = V^{-1},\ee
where $V=\int_\Lambda d\lambda$ is the volume of the domain.

We now show how to calculate $p_e(\lambda)$, which is the output of the Bayesian estimation. This can be achieved by calculating $\tilde{p}_e(\lambda)$ and integrating it (numerically), and then substituting in \eq{Eq:Bayes1}. The likelihood of an experiment consisting of $n_s$ sequences, each repeated $n$ times (shots), is given by the binomial distribution
\be p(\{m\}_e|\lambda)\propto \prod_k^{n_s} (p_k)^{s_k}(1-p_k)^{n-s_k},\label{Eq:Binomialpropto}\ee
where $p_k$ is the model-predicted probability (given the parameters) and $s_k$ the count of successes, and the binomial coefficients (which do not depend on the parameters), have been omitted. The likelihood $p(\{m\}_e|\lambda)$ can be calculated from the log-likelihood defined by 
\be L \equiv \ln p(\{m\}_e|\lambda),\label{Eq:log_L_def}\ee
 which reads explicitly
\be L = \sum_k\left[ s_k\ln p_k + ({n-s_k}) \ln(1-p_k)\right] +C,\label{Eq:log_L}\ee
where $C$ is an unimportant additive term that can be dropped. Exponentiating $L$ to get $p(\{m\}_e|\lambda)$ and plugging it in \eq{Eq:Bayes0} allows one to get the unnormalized posterior distribution $\tilde{p}_e(\lambda)$.

A Monte Carlo estimate of the integral of an arbitrary function $f(\lambda)$ over the domain $\Lambda$,
\be F= \int_\Lambda f(\lambda)d\lambda,\ee
 can be approximated using $n$ sample points $\lambda_i$ drawn according to a uniform distribution over $\Lambda$, by
\be \hat F= \frac{1}{n} \sum_i \frac{ f(\lambda_i)}{V^{-1}}.\ee
With this formula we can integrate  $\tilde{p}_e(\lambda)$ numerically and get $p_e(\lambda)$, the posterior distribution on a sample of points in the parameter space. The numerical challenge of the Bayesian method lies here -- it is necessary to sample the parameter space with enough resolution to resolve the features of the posterior density, which could be concentrated in a small parameter region, or in some cases can even be multi-modal. We discuss this point again in the following sections.

Integrating over $\lambda$ using the distribution in \eq{Eq:Bayes1} gives a mean estimator,
\be \hat\lambda = \int \lambda p_e(\lambda)d\lambda,\label{Eq:Meanestimator}\ee
which is by construction the expected value of the model parameters given the experiment results.
A similar integration over bilinear combination in $\lambda$ gives the covariance matrix indicating error estimates, given by its entries,
\be {\rm Cov}(\hat\lambda) = \int (\lambda^\alpha - \hat\lambda^\alpha) (\lambda^\beta - \hat\lambda^\beta) p_e(\lambda)d\lambda.\label{Eq:Covestimator}\ee
The diagonal elements of the covariance matrix are the variances of each parameter, and the off-diagonal matrix elements express the correlations. In the current work we find it sufficient to study the variances, which already manifest the dependence of each parameter's uncertainty on the other parameters. The covariances and higher moments can be studied in a similar way.

\section{Results from experimental Bayesian estimation}\label{Sec:BayesianResults}

\subsection{SPAM error results}\label{Sec:SPAMResults}

\begin{figure}[t!]
\includegraphics[width=3.4in]{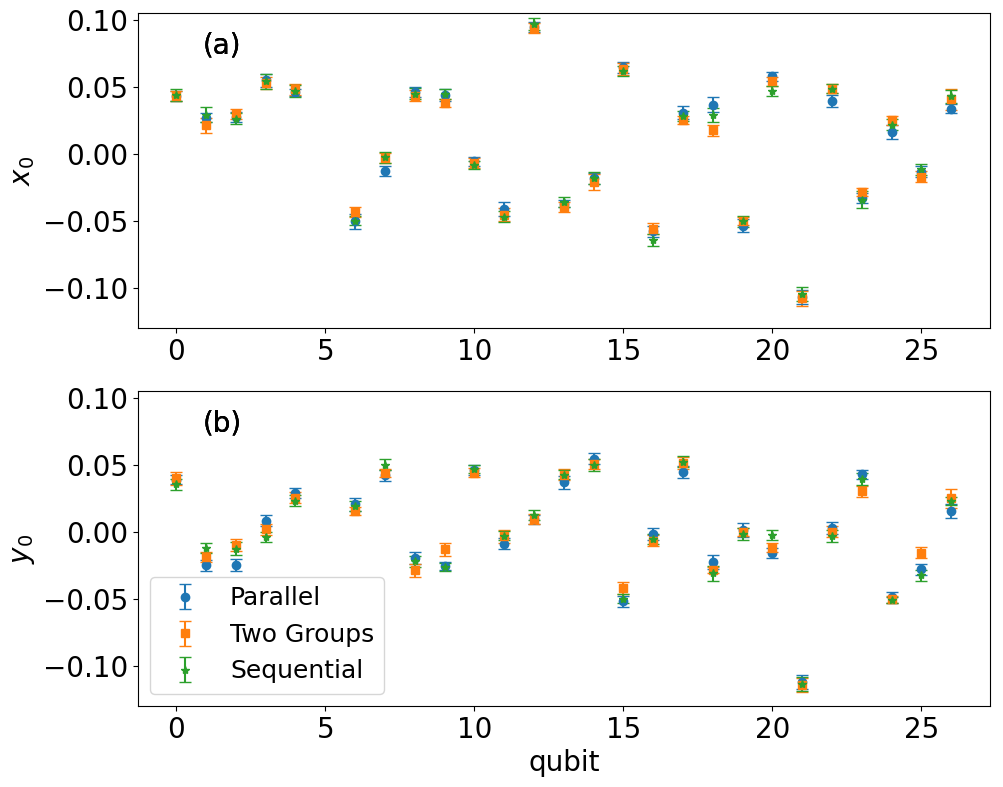}
\caption{Two components of the quantum state preparation (initialization) errors; (a) $x_0$ and (b) $y_0$, as measured using three parallelization modes in one single job on the IBM Quantum device \emph{ibmq\_paris}, accessed via the cloud. 
The estimation method was Bayesian as described in the text (and in \seq{Sec:BayesianIntegrals} in detail), with five state preparation and measurement parameters [\eq{Eq:params}] estimated self-consistently. 
As indicated in the legend, the parameters are estimated either for all qubits in parallel, in two parallel experiments of physically disconnected qubits, or completely sequentially (where only one qubit is being operated on during a given experiment). 
The data is consistent with no clear correlations in the measurement results between the qubits. The error bars are calculated within the Bayesian approach as in \eq{Eq:Covestimator}. Qubit 5 was removed from the analysis because of excessive errors (possibly due to interaction with a transient two-level system in the environment).} \label{Fig:device1}
\end{figure}

In this subsection we present results 
 from the self-consistent estimation of the five SPAM parameters
\be \{x_0, y_0, z_0, \pi_0, \pi_z\}.\label{Eq:params}\ee
Hence, as already discussed above, the following results are based on neglecting gate errors of different types.
With this assumption, the probabilities of measuring 0 in each of the six measurements corresponding to the gates in \eq{Eq:Calibration0}, are respectively given by 
\be f_{\pm z} = \pi_0 \pm \pi_z z_0,\,\,\, f_{\pm y} = \pi_0 \mp \pi_z y_0,\,\,\, f_{\pm x} = \pi_0 \mp\pi_z x_0,\label{Eq:fparams}\ee
which are obtained from \eqss{Eq:fpmz}{Eq:fpmx} by setting $\pi_x=\pi_y=0$.
Choosing any five of the values in \eqs{Eq:fparams} is sufficient for an estimation of the five SPAM parameters but we use all six probabilities for the estimations presented in \figs{Fig:device1}{Fig:mitigation1} of the current and next subsection. Each of the six experiments was measured for 16384 shots. Between $5\times 10^6$ and $80\times 10^6$ points in the parameter space have been sampled for the Bayesian estimation (as described in \seq{Sec:BayesianIntegrals}), which results in a very precise sampling in the current experiments, and is highly efficient with the analytic expressions being used.

\begin{figure}[t!]
\includegraphics[width=3.3in]{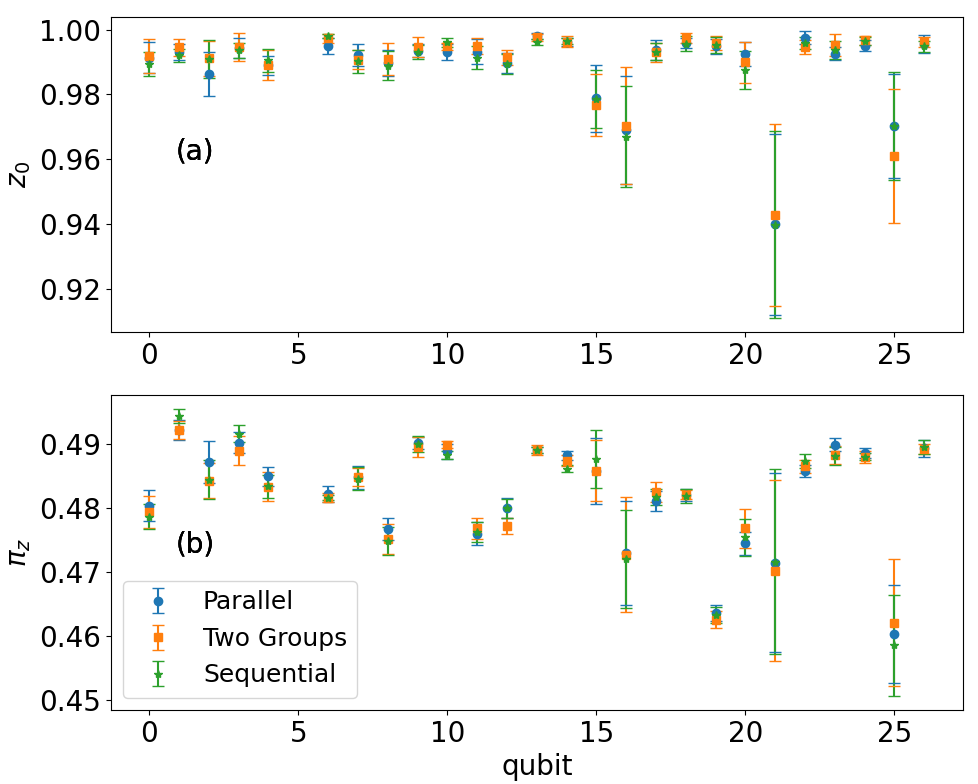}
\caption{(a) The initial state component $z_0$ and (b) one POVM parameter corresponding to classical measurement errors, $\pi_z$, as estimated in the same experiment of \fig{Fig:device1}. For an ideal readout without errors, $\pi_z=0.5$. Here $\pi_z<0.5$ for all qubits, which indicates that measuring 0 when the qubit is in $|1\rangle$ is slightly more probable than the reverse error, consistent with $T_1$ decay events during measurement. See the text for a discussion of the correlation in the error bars seen for some of the qubits, between the classical state preparation and readout error parameters.} \label{Fig:device2}
\end{figure}

Starting with an experiment quantifying the correlations between different qubits, we present the results of one job (a large experiment run contiguously on the device) composed of three different sub-experiments on the IBM Quantum device \emph{ibmq\_paris} \footnote{The experiment was run on the device \emph{ibmq\_paris} on 05/07/2021, with backend version 1.7.16. Backends are listed in https://quantum-computing.ibm.com/}. Figures \ref{Fig:device1}-\ref{Fig:device2} present $x_0,y_0$ and $z_0, \pi_z$,
where the three characterization modes were: (i) all qubits in parallel, (ii) the qubits divided into two groups of disconnected qubits, with the qubits in each group measured in parallel (but sequentially with the other group), and (iii) all qubits measured sequentially. 
The near-coincidence (within error bars) of the three types of experiments for almost all parameters and qubits indicates that to a large extent these parameters are uncorrelated. It is beyond the scope of the current work to quantitatively study small residual correlations that may nevertheless be present. 

\begin{figure}[t!]
\includegraphics[width=3.4in]{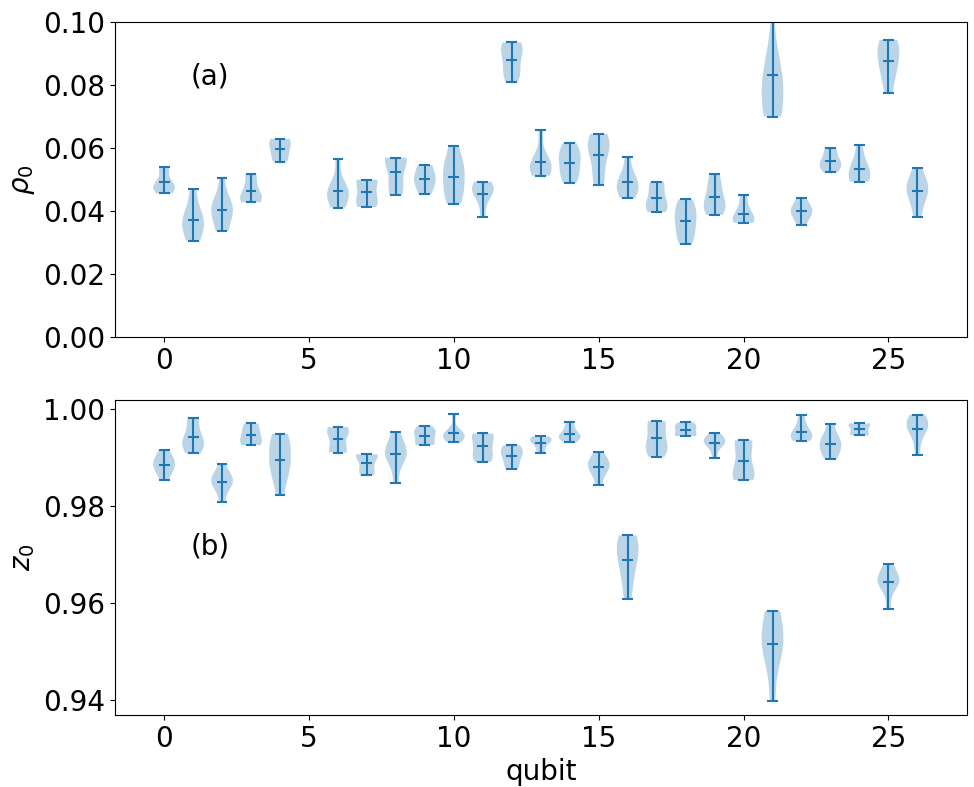}
\caption{(a) The transverse [$\rho_0$ of \eq{Eq:rho_0}] and (b) the initial longitudinal ($z_0$) components of each qubit in the 27-qubit IBM Quantum device \emph{ibmq\_paris}, accessed via the cloud. Seven identical experiments were run at different times within a window of 8 hours. The plot depicts with horizontal bars the minimal, mean and maximal values estimated for each qubit, while the ``violin'' shape indicates the averaged probability shape (being narrow near less frequently observed values). Typical quantum initialization errors ($\rho_0\sim 0.05$) are larger than the classical errors (given by $1-z_0\sim 0.01$), and the initialization errors are mostly relatively stable for time scales of hours. Qubit 5 was removed from the analysis because of excessive measurement errors.} \label{Fig:device0}
\end{figure}

We emphasize an important useful characteristic of Bayesian estimation that manifests itself in the presented figures, where the error bars already account for both the statistical uncertainty and the effect of the self-consistently estimated readout errors. Indeed in \fig{Fig:device2} the correlation between $z_0$ and $\pi_z$ can be clearly seen for a few qubits, where larger deviations of $z_0$ from the ideal value of $1$, lead to larger estimation errors of both of these initialization and readout parameters, and such correlations between various parameters can be systematically analyzed with the current approach. The constraint $\pi_0+\pi_z\le 1$ (expressing that the probability of measuring any result is bounded by 1) is enforced on the domain of the parameter distribution, and correspondingly $\pi_0$ (not shown) takes values slightly larger than 0.5 for all qubits. It should be noted that qubit 5 had noticeably worse measurement parameters ($\pi_z,\pi_0$) possibly due to an interaction with two-level systems (TLSs) and we have therefore avoided plotting the corresponding estimated parameters.

As can be seen in \fig{Fig:device1}, the quantum state preparation error appears to take different orientations in $x-y$ plane for each qubit. A natural question that arises is whether this coherent initialization error fluctuates or drifts between experiments, and on what timescales it can be considered stable.
Figure~\ref{Fig:device0} presents a characterization of the qubits of the IBM Quantum device \emph{ibmq\_paris} \footnote{The experiments were run on the device \emph{ibmq\_paris} on 04/28/2021, with backend version 1.7.15.}, as measured in seven identical experiments conducted over the course of eight hours (the duration of each experiment was approximately 1 minute). The figure presents estimates of the initial longitudinal ($z_0$) and transverse ($\rho_0$) components of each qubit, where the polar Bloch vector parametrization takes the obvious definition using the $xy$-plane angle $\varphi_0$,
\be x_0 \equiv\rho_0 \cos\varphi_0,\qquad  y_0 \equiv \rho_0 \sin\varphi_0.\label{Eq:rho_0}\ee

\begin{figure}[t!]
\includegraphics[width=3.4in]{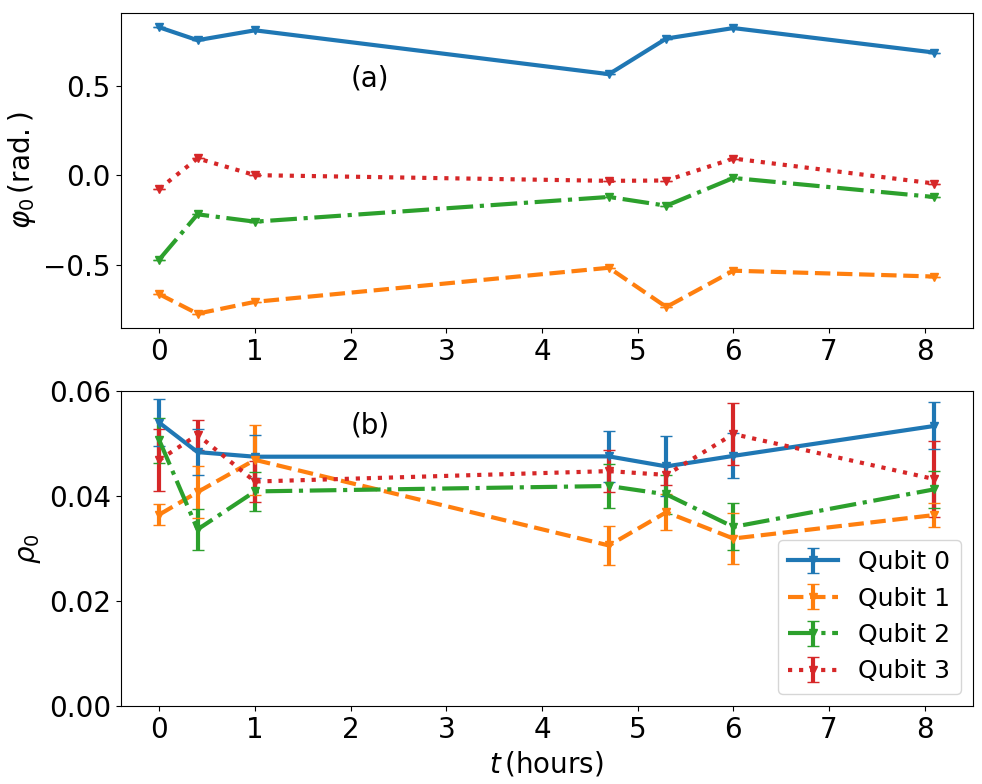}
\caption{(a) The plane polar angle ($\varphi_0$),  and (b) the initial transverse component ($\rho_0$),  both defined in \eq{Eq:rho_0}, of the first four qubits (chosen arbitrarily), plotted vs.~time for the same experiments as in \fig{Fig:device0}. Seven identical experiments were run at different times within a window of 8 hours. The initialization errors can be seen to be relatively stable for time scales of hours.} \label{Fig:device-time}
\end{figure}

The data in \fig{Fig:device0} clearly shows that the typical quantum initialization errors ($\rho_0\sim 0.05$) are larger than the classical errors (given by $1-z_0\sim 0.01$). Moreover -- with a few exceptions, the initialization errors are relatively stable on the time scale of hours, which can be seen also for the qubits' plane polar angles ($\varphi_0$) in the curved of \fig{Fig:device-time}. We study the consequences of this observation in the following subsection. In \seq{Sec:GateResults} we also study gate error parameters, finding that their contribution is typically (in the studied devices) of the order of the error bars in the estimation of the SPAM parameters (or less), which justifies \emph{a-posteriori} their neglect in the above treatment.

\subsection{SPAM mitigation results}\label{Sec:SPAMMitigationResults}

The observed stability of the quantum initialization data suggests that the characterization of these coherent errors can be corrected {\it a-priori}. There are two possible ways for such a correction -- either the state preparation can be calibrated to account for and reduce such errors, or the characterization data can be used to perform a pre-rotation of each qubit at the beginning of a subsequent experiment. The calibration of the state preparation is beyond the scope of the current work because not all users of a quantum device have the capability of doing calibrations when using a device via cloud access. We therefore focus here on the second possibility of rotating the qubits. In \seq{Sec:Outlook} we briefly discuss mitigation when the error parameters fluctuate on a timescale that does not allow to separately characterize and then rotate the qubits. In such a scenario, when the errors are still coherent over the duration of one experiment, they can be characterized in parallel with the execution of each particular experiment, and then mitigated {\it a-posteriori} by post-processing the experiment results together with the characterization data. We will not follow this approach in the current work.

\begin{figure}
\includegraphics[width=3.3in]{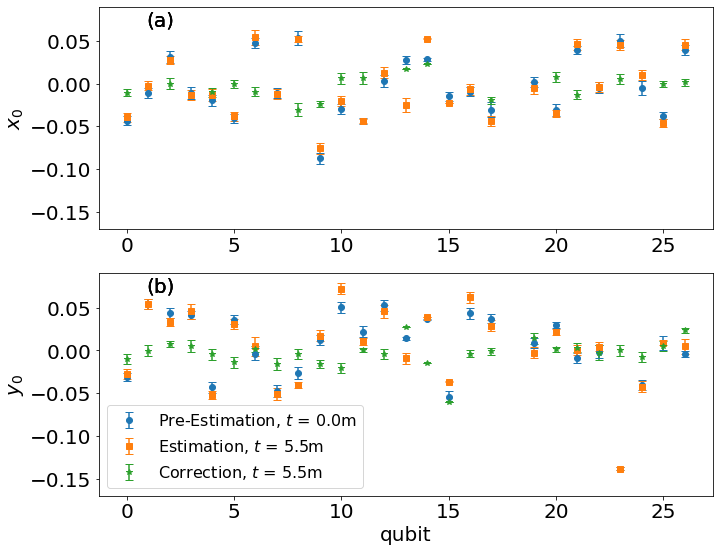}
\caption{Two components of the quantum state preparation errors, (a) $x_0$ and (b) $y_0$, for each qubit in the 27-qubit IBM Quantum device \emph{ibmq\_toronto}, in an experiment of estimation and mitigation run in two steps.  The first estimation starting (arbitrarily) at $t=0$,  was used to calculate the parameters needed for mitigation of the state preparation errors.  The second experiment included two estimations of $x_0$ and $y_0$ parameters,  with and without their mitigation by rotation gates.  As can be seen, the mitigation reduced the initialization error in the device. Here, qubit 18 was removed from the analysis because of excessive errors. See the text for a detailed description of the protocol employed.} \label{Fig:init_error_mitigation}
\end{figure}

\begin{figure}
\includegraphics[width=3.3in]{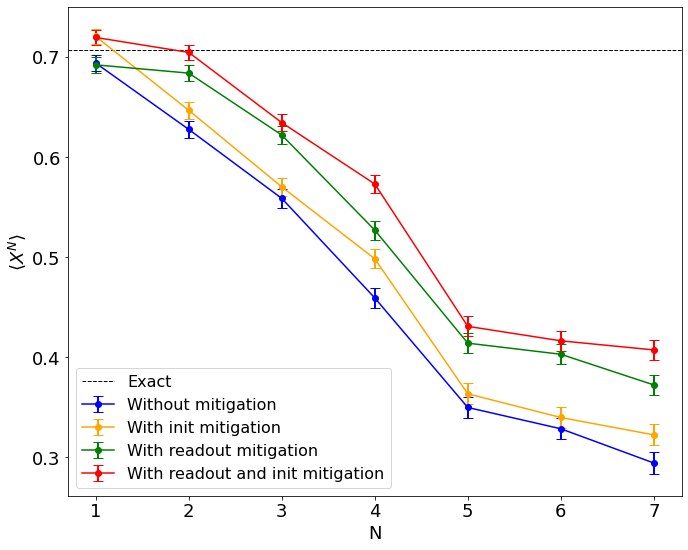}
\caption{Comparison of expectation values obtained using different possible steps of mitigation, in an experiment performed on the 27-qubit IBM Quantum device \emph{ibmq\_sydney}, accessed via the cloud. In each experiment the $N$-qubit cat state of \eq{Eq:psi_cat} was prepared (with some inevitable noise), and then the expectation value of the $N$-qubit operator $\otimes X^{N}$ was measured,  while $N$ was increased from step to step. Each expectation value was calculated four times; without any mitigation, with pre-rotation of the participating qubits in order to mitigate the estimated state preparation errors, with mitigation of the readout errors using the estimated readout error probabilities, and with both state preparation and readout mitigation. The ideal expectation value is shown with a dashed line.
These results show that the estimated SPAM parameters of the device were accurate enough to enable non-negligible improvements of the results via error mitigation.} \label{Fig:mitigation1}
\end{figure}

Figure \ref{Fig:init_error_mitigation} shows the estimation of the quantum state preparation errors,  $x_0$ and $y_0$, both before and following its mitigation, in an experiment run in two steps \footnote{The experiment was run on the device \emph{ibmq\_toronto} on 01/31/2022 with backend version 1.6.9.}. First, an estimation of the device parameters given in \eq{Eq:params} was performed. Using these parameters,  the spherical coordinates of the initial Bloch state vector of each qubit were calculated by considering
\be \theta_0 = \tan^{-1}\left(\rho_0, z_0\right), \ee
in addition to $\rho_0$ and $\varphi_0$ defined in \eq{Eq:rho_0}. In the second step of the experiment, the mitigation of the initial $xy$-plane component was done using two consecutive rotation gates; An $R_z$ gate (a single-qubit rotation about the z axis) with a rotation angle of $-\varphi_0$ was first applied in order to align the initial state vector with the $x$ axis, and an $R_y$ gate (a single-qubit rotation about the y axis) with a rotation angle of $-\theta_0$ was applied in order to align the initial state vector with the $z$ axis. 

Together with the mitigation gates in the second step, we estimated the state preparation errors of the device again, both with the additional rotation gates and without them, in order to test the mitigation process. As can be seen in \fig{Fig:init_error_mitigation}, for almost all qubits the state preparation errors were reduced significantly with the mitigation. By comparing the parameters of the pre-estimation and the estimation without mitigation, drifts of these parameters can be observed for few qubits. These drifts caused the mitigation on those qubits to be less effective. One can also notice that the mitigation of the $y_0$ parameter tends to be better. That is plausibly due to the $R_z$ gate used to mitigate the $y_0$ error being a virtual gate without any error, as opposed to the $R_y$ gate which mitigates the $x_0$ error and entails an $X_{90}$ rotation basis gate. To avoid the introduction of spurious noise, we avoid mitigating the $x_0$ error with qubits for which $|x_0|<0.015$, which is the reason for some mitigation points missing in the top panel of \fig{Fig:init_error_mitigation}.

As a further independent demonstration of the utility of the mitigation of SPAM errors,, \fig{Fig:mitigation1} shows the results of a simple quantum computing experiment \footnote{The experiment was run on the device \emph{ibmq\_sydney} on 08/05/2021 with backend version 1.0.68.}, wherein a circuit for generating a multi-qubit entangled GHZ (cat) state  with $N=1-7$ qubits was first executed. The circuit produces the state 
\be |\psi\rangle\propto \left(|0\rangle^{N} + e^{i{\pi}/{4}}|1\rangle^{N}\right),\label{Eq:psi_cat}\ee 
(which in practice was created with some inevitable noise), and then the expectation value of the global operator $X^{N}$ was measured \footnote{The phase factor was chosen to avoid the expectation value being 0 or 1, which are nongeneric since they are at the boundary.}. The parameters obtained in a preceding estimation experiment were used in order to perform a pre-rotation of the qubits as described above. In order to mitigate the readout errors, the readout parameters presented in \eq{Eq:params} were used to calculate $\epsilon_{0}$ and $\epsilon_{1}$ using \eqss{Eq:epsilon0}{Eq:epsilon1}. A matrix of transition probabilities was calculated for each qubit based on these results (while assuming that the readout noise of each qubit is independent of the other qubits).
With these matrices, the mitigated readout results were estimated using least-squares fitting. For simplicity, in the estimation of SPAM parameters we used here we neglected the gate errors both in the estimation circuits and in the state initialization mitigation (the rotation gates). These gate errors are smaller than the SPAM errors for most qubits, but might cause over mitigation, especially with a low number of qubits. Nevertheless, the experiment shows that both state preparation and readout errors of a device can be estimated and mitigated using the proposed methods.

\subsection{Gate error results}\label{Sec:GateResults}

\begin{figure}
\includegraphics[width=3.4in]{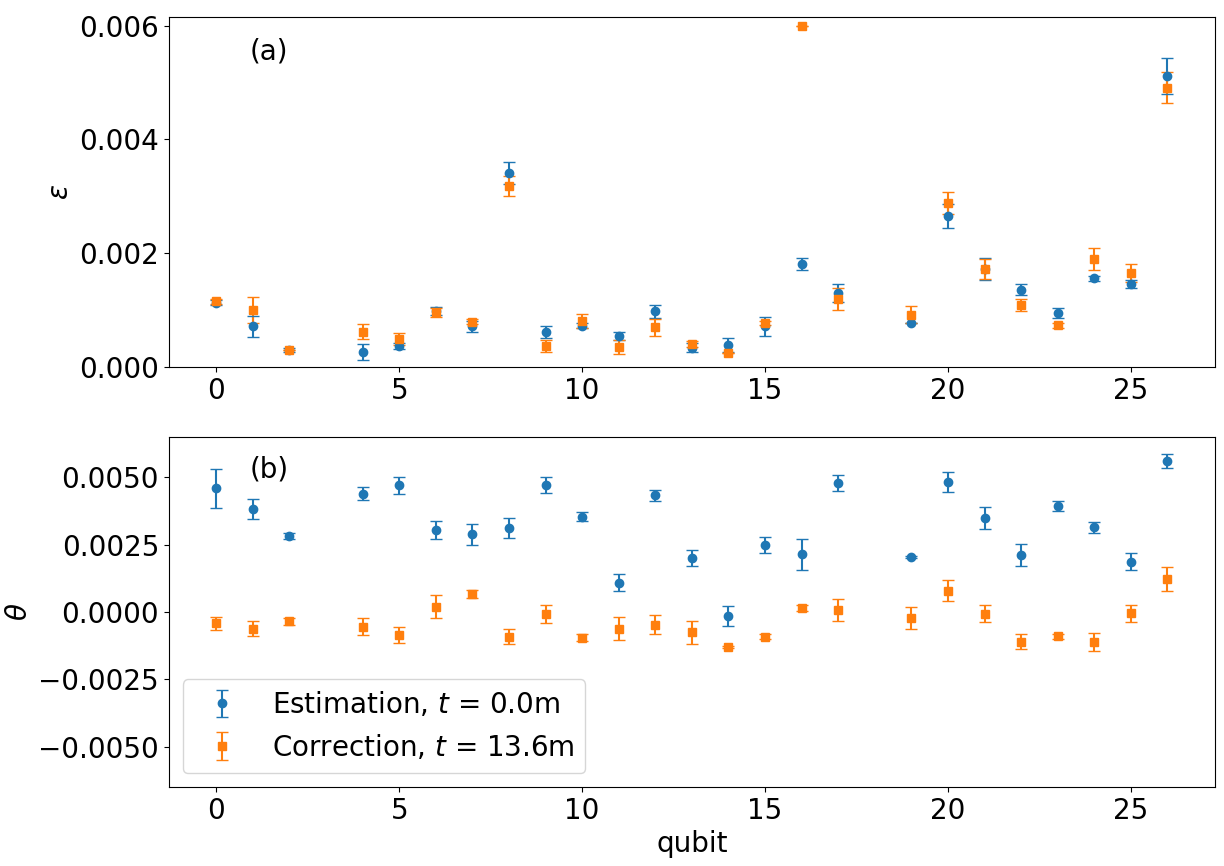}
\caption{(a) The gate depolarization error parameter [$\varepsilon$ of \eq{Eq:depolarizing}], and (b) gate rotation error parameter [$\theta$ of \eq{Eq:X90rotation}], of each qubit in the 27-qubit IBM Quantum device \emph{ibmq\_sydney}. As indicated in the legend, two experiments are shown, with the first experiment starting (arbitrarily) at $t=0$. For the second experiment, a small heuristic correction to the gate pulse amplitude was applied based on the first experiment's estimated rotation error (see the text for more details). As can be seen, this \emph{ad hoc} correction has significantly reduced the overrotation error for most qubits. Here, the Bayesian estimation method described in the text, with seven state preparation, measurement and gate error parameters were self-consistently estimated using eight gate sequences [\eq{Eq:Calibration1}].
Here, qubits 3 and 18 were removed from the analysis because of excessive measurement errors.} \label{Fig:gate-correction}
\end{figure}
 
 In order to estimate the two single-qubit gate error parameters together with the SPAM errors, we repeatedly apply the $X_{90}$ gate.
 We use eight gate measurements in order to self-consistently estimate the seven parameters in Eqs.\eqref{Eq:params}-\eqref{Eq:X90rotation} and \eqref{Eq:depolarizing}, with the gates in \eq{Eq:Calibration0} extended to read
 \be \left\{\mathds{1},X,X_{-90},X_{90},Y_{90},Y_{-90},(X_{90})^{32}, (X_{90})^{33}\right\}.\label{Eq:Calibration1}\ee
 As in \seq{Sec:SPAMResults}, each gate sequence was measured for 16384 shots in the experiments described below, and $80\times 10^6$ points in the parameter space have been sampled for the Bayesian estimation.
  
Figure \ref{Fig:gate-correction} shows the results of such an estimation on the 27 qubits of the IBM Quantum device \emph{ibmq\_sydney} \footnote{The experiment was run on the device \emph{ibmq\_sydney} on 05/20/2021 with backend version 1.0.47.}, as measured with two experiments separated by approximately 14 minutes. The figure presents estimates of $\varepsilon$ and $\theta$ for each qubit. Following the first estimation (at the arbitrary time $t=0$), the estimated rotation errors are then corrected by adjusting the gate pulse amplitude using the Qiskit Pulse interface \cite{alexander2020qiskit}, which allows us to control parameters of the microwave pulses sent to the qubits. When the gate error is small, the correction is expected to be linear, and the relation that has been used for \fig{Fig:gate-correction} reads
\be A\to (1-\eta \frac{\theta}{\pi})A, \qquad \eta=1.33\ee
where $A$ is the amplitude of the $X_{90}$ microwave pulse envelope (which is a square pulse with a Gaussian-shaped rise and fall on both sides), and we have heuristically set the value of the proportionality coefficient $\eta$ to 1.33.
As \fig{Fig:gate-correction} shows, in this example, this heuristic correction has significantly reduced the rotation error for most qubits, even though it was applied almost a quarter of an hour after the estimation experiment.

\begin{figure}
\includegraphics[width=3.4in]{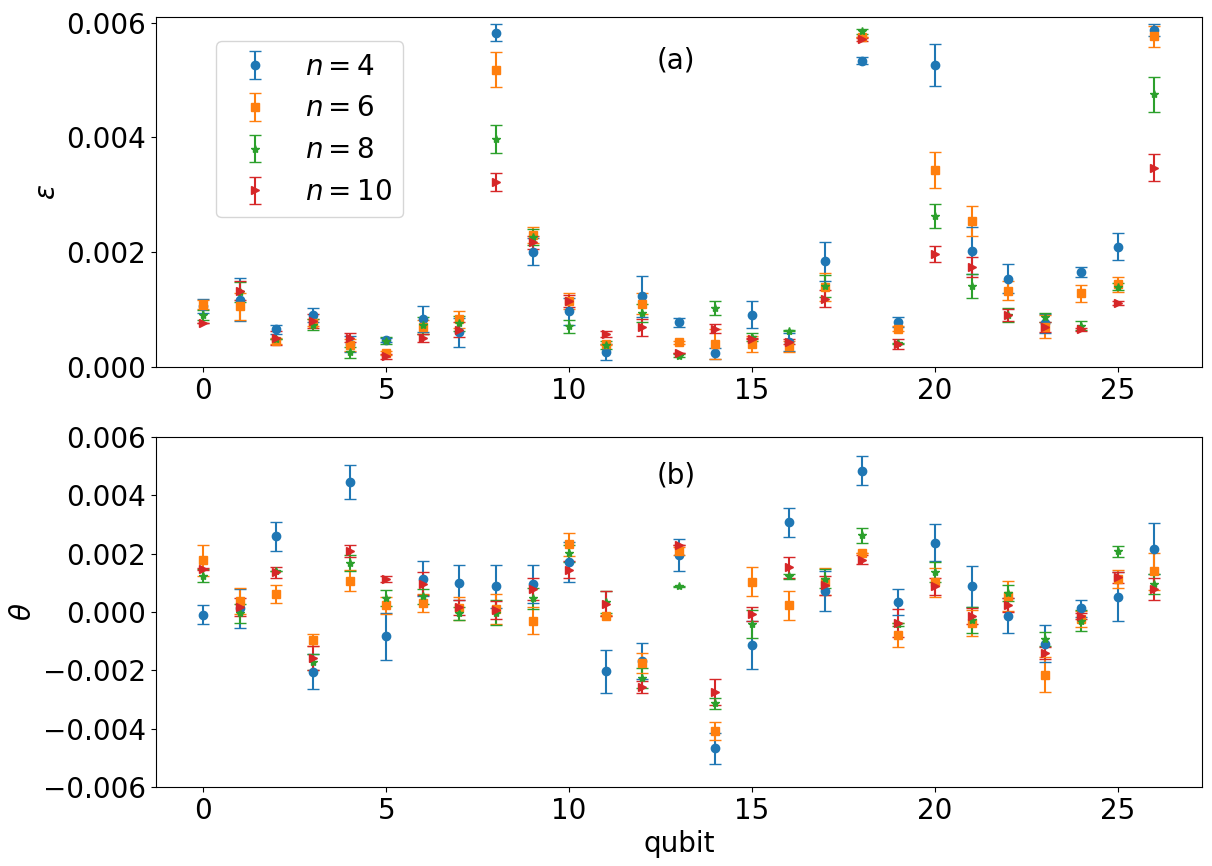}
\caption{The two single-qubit gate error parameters of each qubit in the 27-qubit IBM Quantum device \emph{ibmq\_sydney}, as a function of the powers that the $X_{90}$ gate is taken to, with $n$ being the exponent in $(X_{90})^{4n}$ and $(X_{90})^{4n+1}$. 
It can be seen that for these parameters, the estimated $\theta$ values typically converge for $n\gtrsim 6$, and the same holds for $\varepsilon$, with a few exceptions. 
} \label{Fig:gate-power}
\end{figure}

In \fig{Fig:gate-power} we present data \footnote{The experiment was run on the device \emph{ibmq\_sydney} on 05/24/2021 with backend version 1.0.47.} comparing the estimation of the two gate error parameters $\varepsilon$ and $\theta$ as function of the number of repetitions of the $X_{90}$ gate in the set of gates parametrized using the integer $n$ as
 \be \left\{\mathds{1},X,X_{-90},X_{90},Y_{90},Y_{-90},(X_{90})^{4n}, (X_{90})^{4n+1}\right\}.\label{Eq:Calibration8}\ee
The convergence and accuracy of the estimation depends on $n$ and the values of the parameters themselves, and here $\theta$ values noticeably converge for $n\gtrsim 6$, while there appears to be more variability in $\varepsilon$. Comparing the $\theta$ data in \fig{Fig:gate-correction} to \fig{Fig:gate-power}, it can be seen that in the former experiment  $\theta\ge 0$ for all qubits, while in the latter, the estimated $\theta$ values (for $n=8,10$, which are reliably converged) are smaller in absolute values, and also can be seen to take both positive and negative value. This could indicate a correlated fluctuation in the parameters affecting the device during in the experiment of \fig{Fig:gate-correction}.

\begin{figure}[!t]
\includegraphics[width=3.4in]{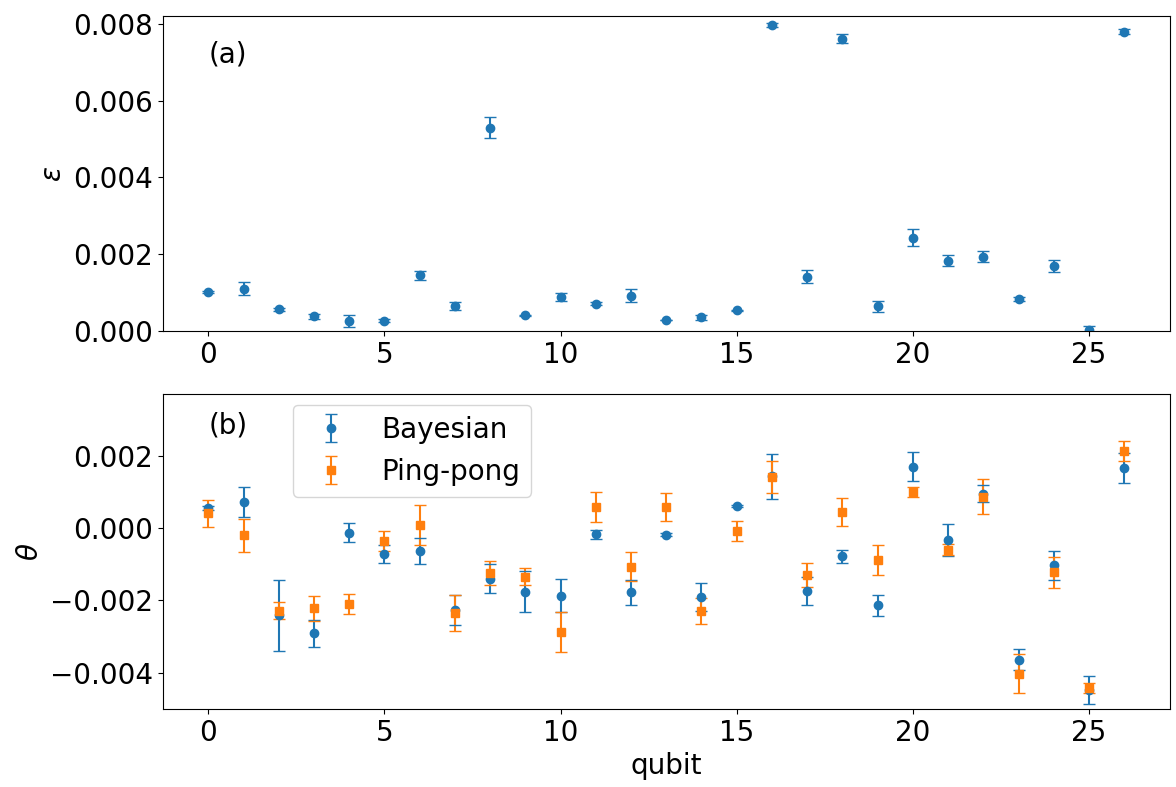}
\caption{Bottom: The ping-pong comparison experiment for all qubits of the 27-qubit IBM Quantum device \emph{ibmq\_sydney}. Top: The depolarization parameter estimated using the Bayesian method in the same experiment, shown for reference. For most qubits, despite a few notable exceptions, the estimation of $\theta$ using the Bayesian approach is close (within statistical fluctuations) to the estimation using an extended ping-pong method with variation of the gate amplitude. This serves as an independent validation of the Bayesian estimation parametrization and experiments.
} \label{Fig:ping-pong}
\end{figure}

As a further corroboration of the accuracy and relevance of the gate error parameters that we estimate, we compare the Bayesian estimation results and the well-established method for calibration and estimation of gate rotation errors, commonly known as the ping-pong calibration \cite{sheldon2016characterizing}. In \app{App:Pingpong} we give more details of the experiment and fitting procedure that we applied, and present an example of the resulting data and plots. Figure \ref{Fig:ping-pong} shows (bottom panel) a comparison of the ping-pong estimates against the Bayesian method for all qubits of the device \footnote{The data was taken on the device \emph{ibmq\_sydney} on 06/28/2021, with backend version 1.0.53.}. With just a few exceptions, the estimated $\theta$ values for most qubits are equivalent within the expected statistical fluctuations as described by the error bars. In the top panel of the same figure, the values of $\varepsilon$ derived in the Bayesian estimation are given for completeness. Beyond being an independent validation of the gate error estimations using the Bayesian approach, the above experiment provides a starting point for a comparison of the use of quantum resources (experiment durations and number of shots) between different approaches. While a more extensive such analysis is beyond the scope of the current work, in \seq{Sec:Outlook} we discuss in more detail some initial conclusions that the above data suggests.

Finally, we comment on the question of the reliability of the $\varepsilon$ parameter estimations. The duration of the single qubit gates in those experiments is approximately $35\,$ns. The $T_1$ (energy decay timescale) and $T_2$ (decoherence timescale) values reported by the device for the qubits (e.g., in the experiment of \fig{Fig:gate-correction}), are in the range of $30-180\,\mu$s, with most being around $70-100\,\mu$s. Hence the shortening of the Bloch vector corresponding to one gate execution, as parametrized by $\varepsilon$, can be estimated to be of order 0.0005, with some qubits being worse or better -- matching well our experiment results. The averaging of the action of $T_1$ and $T_2$ dynamics to an effective depolarizing error during the gate could be considered as a zeroth order approximation, and improving it would require a significant complication of the model. We note that if the model is too different from the actual system, it can be expected that the data just represents noise when being fit to the model, or the parameters take values on the boundaries of the priors, and this would be evident in the results. We did not observe such red flags (except for some qubits that were badly performing), and in contrast, at least in some cases as our results show, this approximation of a depolarizing error is still enough in order to facilitate the method capturing the coherent overrotation error to a useful degree.

\section{Summary and outlook}\label{Sec:Outlook}

Using a Bayesian parameter estimation approach with a single-qubit error model we were able to identify key characteristics of different device preparation, measurement and gate errors. The primary advantage of the approach is that it allows one to simultaneously and self-consistently estimate many error parameters, utilizing the full information available in a set of experimental data. This advantage comes along with physical constraints being respected {\it a-priori}, and together with accessible estimation errors and covariances. Although we have not pursued this line of research, the available information about the parameter correlations and interdependency can be valuable for specific purposes.

Applying our approach to study qubit errors in the 27-qubit \emph{ibmq\_sydney}, \emph{ibmq\_toronto}, and  \emph{ibmq\_paris} devices we were able to identify non-negligible state preparation errors of a quantum nature. These preparation errors correspond to a small coherent rotation of the ideal initial state following the initialization. As mentioned above, the physical origin of the quantum state preparation errors is rooted in the unconditional reset method, however, for our estimation procedure this fact is irrelevant and the parametrization is agnostic of the underlying mechanism. With the observed stability of this error and the high precision of its estimation, we have demonstrated its mitigation in by adding a pre-rotation of each qubit, compensating for its coherent reset error. Expectation values of a multi-qubit operator were corrected by a significant fraction in the example that we studied, showing the relevance of the characterized errors.

Augmenting the self-consistent SPAM error experiments with two more gate sequences, we demonstrated a high precision estimation of single qubit gate errors, parametrizing one incoherent error (a depolarizing channel), and one coherent rotation error.
Similar to the coherent quantum state preparation errors, the gate rotation error lends itself to immediate mitigation (in an experiment following its estimation), using the cloud based interface for pulse control of qubit devices.
These two approaches (qubit pre-rotation and gate amplitude scaling) to correct the coherent quantum errors can be considered as ``online'' mitigation which is a simplified (and straightforward) form of calibration based on using the error estimates obtained in one experiment in order to reduce the errors in the next experiment.

In addition, one can consider ``offline'' mitigation which is performed via post-processing of measurement results. This approach is suitable when the error parameters fluctuate on a timescale that does not allow separate characterization and correction of errors before running a computation (or when this is infeasible for other reasons). In such a scenario, when the errors are still coherent over the duration of one experiment, they can be characterized in parallel with the execution of each particular experiment, and then mitigated {\it a-posteriori} by post-processing the results together with the characterization data. Readout errors are the primary (and widely used) example of this approach.
Also, mitigation and state and process tomography codes (e.g., those included in Qiskit \cite{Qiskit}), can be initialized with error parameters such as the quantum errors observed in the current work. The source code facilitating estimation experiments such as we have analyzed in this work \cite{quantum-bayesian-estimation-repo} can be used together with Qiskit to extract these parameters.

We conclude by commenting on the important aspect of the scalability and resources utilized by the experiment. The Bayesian estimation experiments presented in \seq{Sec:GateResults} consisted of an application of the 8 gate sequences (of different lengths, from single-gates and up to 33 gates) listed in \eq{Eq:Calibration1}, from which 7 parameters that were estimated. In a characterization of the gate rotation error parameter $\theta$ performed to compare to an extended ping-pong method (as shown in \fig{Fig:ping-pong}, and in more detail in \app{App:Pingpong}), 80 gate sequences (of varying lengths as well) were necessary for the estimation of just this one parameter with approximately the same accuracy.

 While a more complete exploration of gate calibrations and the required resources using the Bayesian approach are beyond the scope of the current work, this demonstrates the potential power of the Bayesian method in estimating small parameters with a high precision, while significantly reducing the number of measurements required on the quantum hardware. When considering interacting qubits (such as for calibration of two-qubit gates), and the number of parameters and the complexity of the model grow, the space to be explored in Bayesian estimation increases in dimension, making it hard to estimate all parameters together. However, it is not necessary in general to estimate {\it all} model parameters together; Well chosen parametrizations of the errors and the experiments can be beneficial in coping with this issue by splitting the estimation into nearly-independent groups of parameters. Pushing calibrations to a high accuracy while maintaining a low exploitation of quantum resources is an important goal in the field, which may prove important for reaching the low error thresholds required for fault-tolerant quantum computing.

\begin{acknowledgments}
We thank Gadi Aleksandrowicz, Shelly Garion, Eli Arbel, and Sam Ackerman for very helpful feedback. We thank Nick Bronn, Isaac Lauer, Doug McClure, Matthias Steffen and Oliver Dial for fruitful discussions about the unconditional reset mechanism and the qubit preparation errors observed in the experiments.
\end{acknowledgments}

\appendix

\section{Characterization using maximum-likelihood estimators of POVM errors}\label{Sec:POVM4}

First, let us write the positivity conditions on the POVM parameters of \seq{Sec:POVMErrors}.
Assuming that 
\be 0 \le \epsilon_0 < 1, \qquad 0 \le \epsilon_1 <1,\ee
 the positive semi-definiteness of both $ \Pi $ and  $ \mathds{1}-\Pi $ gives us the conditions (using the determinant),
\be (\pi_{0})^2,\, (1-\pi_{0})^2 \ge \sum_{a\in\{x,y,z\}} (\pi_{a} )^2, \label{Eq:Piconstraint}\ee
or, equivalently,
\be  \epsilon_1(1-\epsilon_0),\, \epsilon_0(1-\epsilon_1)\ge ( \pi_x^2+ \pi_y^2).\ee

If the initialization and single-qubit gate errors can be neglected, the estimation of the distribution of the four POVM parameters $\pi_\mu$ of \eq{Eq:Pi} can be done directly. The characterization in this case is straightforward due to the linear dependence of the measurement probabilities on the parameters, and the Bayesian numerical procedure is not required. The experiment is composed of four independent measurements, each preceded by one of the gates in the set (other choices could suitable as well),
\be \left\{\mathds{1},X,X_{-90},Y_{90}\right\}.\label{Eq:Calibration4}\ee

Under the assumption that the initial state is $|0\rangle$, the probabilities of measuring 0 in each of the four calibration experiments using the gates in \eq{Eq:Calibration4} are given, respectively, by 
\be f_{z} = \pi_0+\pi_z, \qquad f_{-z} = \pi_0-\pi_z,\ee
\be f_y = \pi_0+\pi_y, \qquad f_x = \pi_0+\pi_x.\ee
The maximum likelihood estimators $\hat{\pi}_\mu$ follow directly from the estimators $\hat{f}_\mu$ from the count of 0 shots in each experiment [as in \eq{Eq:hat_p_binomial}], using the inverse relations 
\begin{alignat}{3}
\hat\pi_0 &= \frac{1}{2}(\hat{f}_{z}+\hat{f}_{-z}), & \qquad \hat{\pi}_z &= \frac{1}{2}(\hat{f}_z -\hat{f}_{-z}),\\
\hat{\pi}_x &= \hat{f}_x - \hat{\pi}_0, & \qquad \hat{\pi}_y &= \hat{f}_y - \hat\pi_0.
\end{alignat}
Using these expressions, we can derive the distribution of the estimators $\pi_\mu$. The analytic derivation of their distribution gives at once the variances of the estimators, indicating the error bars on the POVM parameters.
In the limit of a large number of shots $n$, the binomial distributions from which the experiment samples in each measurement, can be approximated using a Gaussian (normal) distribution. As a rule of thumb, this approximation is taken to be sufficient when the parameters of the distribution of \eq{Eq:Binomial} obey
\be np > q, \qquad n(1-p)>q,\label{Eq:Gaussian_condition}\ee
with $q$ some threshold guaranteeing that both $p$ and $1-p$ are not too small probabilities to skew the Gaussian approximation, and $q=9$ is often chosen, which corresponds to three standard deviations in the normal distribution. Under this assumption, we can write instead of \eq{Eq:Binomial},
\be s_0 \sim \mathcal{N}\left(\mu=np,\sigma^2 =np(1-p)\right),\label{Eq:Gaussian}\ee
i.e., a normal distribution with mean $\mu= np$ and variance $\sigma^2 = np(1-p)$. Hence the estimator $\hat p$ is distributed as
\be \hat p \sim  \mathcal{N}\left(p,\frac{1}{n}p(1-p)\right).\label{Eq:hat_p_Gaussian}\ee

Using the arithmetic of independent normal random variables,
\be\left( a f_1 + b f_2 \right)\sim \mathcal{N}\left(a\mu_1 + b\mu_2,|a|^2 \sigma_1^2 + |b|^2 \sigma_2^2\right),\ee
the distribution of the estimators can be obtained explicitly, e.g.,
\be \hat\pi_{0,z} \sim \mathcal{N}\left(\frac{1}{2}({f}_0\pm {f}_z), \frac{1}{4n}[{f}_0(1-{f}_0)+{f}_z(1-{f}_z)]\right),\ee
and similarly for $\hat\pi_{x}$ and $\hat\pi_{y}$, the distribution is normal with the mean and variance being
\be \mu_{x,y} = {f}_{x,y} - \frac{1}{2}({f}_{z}+ {f}_{-z}),\ee
\be \sigma_{x,y}^2 = \frac{1}{4n}[4{f}_{x,y} ( 1- {f}_{x,y})+{f}_{z} (1-{f}_{z})+{f}_{-z}(1-{f}_{-z})],\ee
 For the values of $f_\mu$ in these formulae, we substitute their estimators obtained in the experiment, $\hat f_\mu$. 
 We therefore have analytic estimates for the variances in the estimation of the unknown parameters $\pi_\mu$ used in the mitigation matrix.
 
 However, it is important to note that the above approximation using normal distributions, we have neglected the constraints on the POVM elements.
The Gaussian distributions above are therefore justified when the conditions in \eq{Eq:Gaussian_condition} hold, and in addition when the constraints of \eq{Eq:Piconstraint} hold, within a few (say, at least three) standard deviations of the normal distributions of the parameters. The Bayesian estimation discussed in \seq{Sec:Bayesian} resolves the issue of constraints in a straightforward manner, and allows the consistent estimation of more parameters.

\section{Explicit overrotations}\label{App:Overrotations}

Using \eqss{Eq:X90rotation}{Eq:Y90rotation} for the case of overrotation parametrized with just one angle $\theta$, we get explicit formulae for a few useful rotation powers,
\be X_{\pm 90}\vec{r} = (x, -\sin\theta y \mp \cos\theta z, \pm\cos\theta y -\sin\theta z), \ee
\be Y_{\pm 90}\vec{r} = (-\sin\theta x \pm \cos\theta z, y, \mp \cos\theta x -\sin\theta z), \ee
\bem (X_{\pm 90})^2\vec{r} =\\ (x, -\cos 2\theta y \pm \sin 2\theta z, \mp\sin 2\theta y -\cos 2\theta z), \end{multline}
\bem (X_{\pm 90})^{4n}\vec{r} =\\ (x, \cos 4n\theta y \mp \sin 4n\theta z, \pm\sin 4n\theta y +\cos 4n\theta z), \end{multline}
 \bem (X_{\pm 90})^{4n+1}\vec{r} = (x, -\sin (4n+1)\theta y \mp \cos (4n+1)\theta z,\\ \pm\cos (4n+1)\theta y -\sin (4n+1)\theta z). \end{multline}

The more general case of an imperfect rotation (deviating from an ideal rotation about the $x$ axis), can be accounted for using a tilted rotation axis  parametrized using two parameters $\hat y, \hat z$ that define the orientation of the axis
\be \hat{u}=\left(\hat x, \hat y, \hat z\right),\qquad \hat x \equiv \sqrt{1-\hat y^2-\hat z^2} \ee
(where the positive root is chosen uniquely when rotations about the $x$ axis are considered), and an angle of rotation $\theta$, giving the rotation matrix
\be X_{\theta} = 1 + (\sin\theta) U + (1-\cos\theta)U^2,\ee
with the generator matrix
\be U =  \left(\begin{array}{ccc}
0 & -\hat z & \hat y \\
 \hat z & 0 &  -\hat x \\
-\hat y & \hat x & 0 \end{array}\right).\label{Eq:X90rotationU}
\ee

\section{The ping-pong estimation experiment}
\label{App:Pingpong}

\begin{figure}
\includegraphics[width=3.4in]{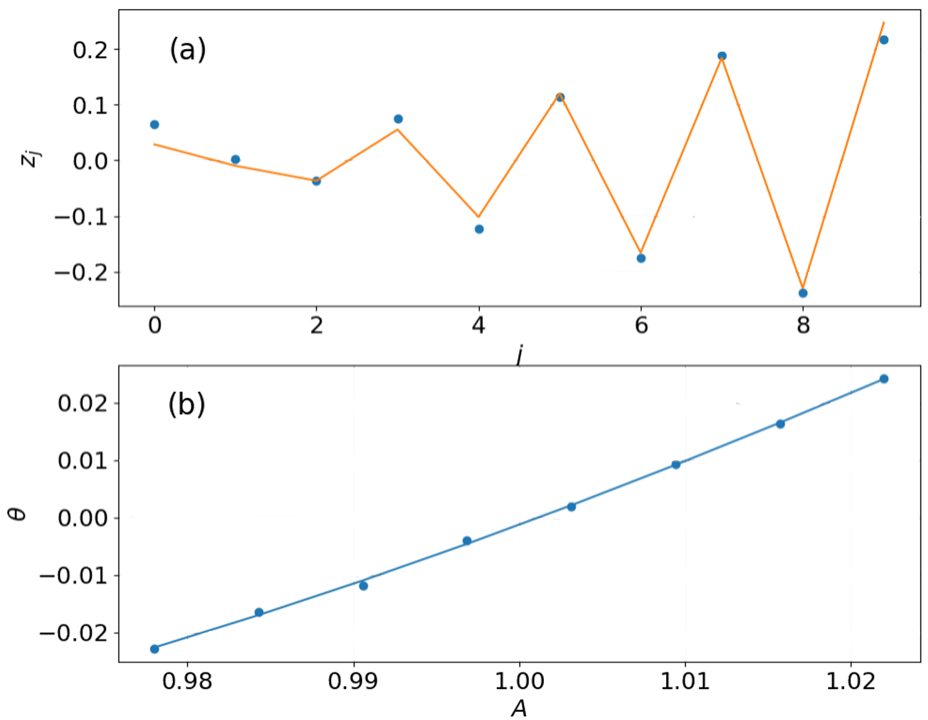}
\caption{The extended ping-pong experiment with a variation of the pulse amplitude, shown corresponding to a single qubit (number 8), for the same experimental run as \fig{Fig:ping-pong-fit}. (a) The experiment points and fit curve of \eq{Eq:z_j} for $A=1.0157$,  and (b) the experiment points and fit curve of \eq{Eq:thetavsA}.
} \label{Fig:ping-pong-fit}
\end{figure}

Precise estimation of a small error on a control parameter requires a technique that can selectively amplify the error of interest. 
A ping-pong technique is one of such experimental techniques that repeats unit sequence to accumulate the error in the form of trigonometric function. 
Here we employ the following sequence to estimate the rotation angle error $\theta$; $X_{+90}-\left[(X_{+90})^2\right]^j$ where $j$ is the number of repetitions of the basic sequence.
The standard ping-pong fitting ansatz is
\be z_j = a+(-1)^j \cos\left(\frac{\pi}{2}+2 j\theta\right).\ee
This form of amplification can be robust to SPAM errors, since such errors are absorbed into offset terms, whereas the rotation angle error appears in the sinusoidal term.
Considering the state preparation error, we extended the ansatz to
\be z_j = a+ (-1)^j\left[b+ \cos\left(\frac{\pi}{2}+2 j\theta\right)\right].\label{Eq:z_j}\ee
When the error $\theta$ is small enough, we can approximate $\left|z_j\right| \sim j\theta$ by ignoring offset terms, namely the error amplification is linear to $j$.
Investigation of calibrated gates when the error is close to zero requires a large number of repetitions, however long experiments usually suffer from the effect of incoherent errors.
Instead, we combine multiple ping-pong experiments varying the amplitude $\Omega = A\Omega_0$ (here $A$ is a scale factor), which allows us to interpolate the value of $\theta$ at $\Omega_0$ (with $A=1$), of the calibrated pulse amplitude.
We fit the resulting curve with a polynomial expansion of the form
\be \theta = \alpha + \beta A + \gamma A^2 + \delta A^3.\label{Eq:thetavsA}\ee
An example of the analysis from such an experiment (with the data used to construct \fig{Fig:ping-pong}), is shown in \fig{Fig:ping-pong-fit} for a single qubit. The top panel shows $z_j$ for a single gate pulse amplitude $A=1.0157$, and the value of $\theta$ extracted from fitting the shown curve, constitutes a single point on the curve in the bottom panel, which is then fitted according to \eq{Eq:thetavsA}, from which the value of $\theta$ at $A=1$ is extracted, with its error bars (that stem from the spread of points of the latter curve).

\bibliographystyle{./hunsrt}
\bibliography{modeling}

\end{document}